\documentclass[
preprint,
prl,
aps,
showpacs,
tightenlines,
amsmath,
amssymb,
]{revtex4} 

\def\grad{\widetilde\nabla}
\def\slash#1{\, /\kern-0.6em{#1}}

\begin{document}

\title{No hair theorems for positive $\Lambda$}     
\author{Sourav Bhattacharya}\email{sbhatt@bose.res.in}
\author{ Amitabha Lahiri}\email{amitabha@bose.res.in} 
\affiliation{S. N. Bose National Centre for Basic Sciences, \\ 
Block JD, Sector III, Salt Lake, Calcutta 700 098, INDIA\\ 
}

\begin{abstract}
We extend all known black hole no-hair theorems to space-times
endowed with a positive cosmological constant $\Lambda.$
Specifically, we prove that static spherical black holes with
$\Lambda>0$ cannot support scalar fields in convex potentials and
Proca-massive vector fields in the region between black hole and
cosmic horizons. We also demonstrate the existence of at least one
type of quantum hair, and of exactly one charged solution for the
Abelian Higgs model. Our method of proof can be applied to
investigate other types of quantum or topological hair on black
holes in the presence of a positive $\Lambda.$

\end{abstract}
\pacs{04.70.Bw}
\maketitle

The no-hair conjecture states that gravitational collapse reaches a
stationary final state, characterized by a small number of
parameters. The part of this that has been rigorously proved,
called the no hair theorem~\cite{Chrusciel:1994sn, Heusler:1998ua,
Bekenstein:1998aw}, deals with the uniqueness of stationary black
holes, which are characterized by mass, angular momentum, and
charges corresponding to long-range gauge fields. In particular,
static black holes do not support external fields corresponding to
scalars in convex potentials, Proca-massive gauge
fields~\cite{Bekenstein:1971hc}, or even gauge fields which have
become massive via the Abelian Higgs mechanism~\cite{Adler:1978dp,
Lahiri:1993vg}.

All these theorems assume, in addition to stationarity, asymptotic
flatness, which requires a vanishing cosmological constant. The
stress-energy tensor then must vanish at infinity, which means that
all matter fields must approach their vacuum values. However,
recent observations suggest a strong possibility that the universe
is equipped with a positive cosmological constant,
$\Lambda>0\,$~\cite{Riess:1998cb, Perlmutter:1998np}. If this is
so, there should be a cosmic horizon of
size~$\sim1/\sqrt\Lambda\,,$ and proofs of uniqueness of black
holes become suspect. Even if a black hole forms as the final state
of gravitational collapse, its horizon will be inside the cosmic
horizon. There is no global timelike Killing vector outside the
black hole horizon.  Further, the stress-energy tensor need not
vanish at infinity, nor even at the cosmic horizon, so boundary
conditions for the fields are not obvious.

Price's theorem~\cite{Price:1971gc}, which may be thought of as a
perturbative no-hair theorem, was proved for $\Lambda> 0$ some
years ago~\cite{Chambers:1994sz} for massless small fluctuations.
But no version of a theorem about the existence of static matter
fields has been established for $\Lambda >0\,.$ Here we establish
classical no-hair theorems for various different fields, and also
extend one known case of quantum hair, on static black hole
space-times with $\Lambda>0\,.$ Our method involves a paradigm
shift -- we consider only the region between the black hole horizon
and the cosmic horizon, and ignore the asymptotic behavior of both
the metric and the matter fields. In fact, we do not use the
equations for the metric at all, beyond assuming the existence of a
cosmic horizon. We find that it is possible to extend most of the
known no-hair theorems to black holes in a universe with
$\Lambda>0\,.$ We also find one clear exception, that of the
Abelian Higgs model.

We will consider the various no-hair conjectures in a black hole
space-time endowed with a positive cosmological constant, which
leads to the existence of a cosmic horizon.  By a static black hole
with $\Lambda> 0$ we will mean a space-time with at least two
horizons, between which there is a timelike Killing vector
$\zeta^{\mu}$ satisfying $\zeta_{[\mu} \nabla_{\nu}\zeta_{\lambda]}
= 0$. Then $\zeta^\mu$ is orthogonal to a spacelike hypersurface
$\Sigma\,,$ which is assumed to be spherically symmetric. The norm
$\lambda(r) = \sqrt{- \zeta^{\mu}\zeta_{\mu}}$ vanishes at two
values $r_{H}< r_{C}$ of the radial coordinate $r\,,$ thus dividing
the manifold into three regions. The region $r< r_H$ contains a
space-time singularity. The points of this region do not lie to the
past $\Sigma$ (for which $r_H < r < r_C\,),$ while the points of
$\Sigma$ do not lie to the past of the region $r > r_C\,.$ We are
not concerned with the world outside the cosmic horizon; so the
asymptotic behaviour of the metric will not be relevant to our
calculations. In particular we do not assume the metric to be
asymptotically de Sitter.

The various no hair theorems will be taken to be statements about
the corresponding classical fields on the spacelike hypersurface
$\Sigma$ between the two horizons. We will not look for solutions,
only prove general statements about their existence. The crucial
ingredient for these proofs is that the squared norm of the
stress-energy tensor is bounded at each horizon. This is dictated
by Einstein's equation, $G_{\mu\nu}=8\pi T_{\mu\nu}-\Lambda
g_{\mu\nu}$; if the stress-energy tensor $T_{\mu\nu}$ has unbounded
norm at any point, the norm of the Einstein tensor $G_{\mu\nu}$
must also become unbounded there, giving rise to a
curvature singularity at that point. Since the horizons are assumed
to be regular, i.e., only coordinate singularities, it follows that
the Einstein tensor and hence the stress-energy tensor must have
bounded norm at both horizons. Similar arguments show that the norm
of the stress-energy tensor must be static, i.e., its Lie
derivative must vanish along the vector field $\zeta^{\mu}$.
Generally we will say that the stress-energy tensor is bounded, or
static, when we actually mean its norm has those properties.

Although the calculations are on a spacelike hypersurface $\Sigma$
orthogonal everywhere to a timelike Killing vector, it is
convenient to use covariant notation without resorting to explicit
coordinates. Let $\Pi^{\mu}_{\ \mu'}=\delta^{\mu}_{\
\mu'}+\lambda^{-2}\zeta^{\mu}\zeta_{\mu'}$ denote the projection
tensor which projects vectors to $\Sigma$ and let
$\widetilde{\nabla}_{\mu}$ denote the induced connection on
$\Sigma$. Then for a rank $p$ antisymmetric tensor $\Omega$ whose
Lie derivative with respect to $\zeta^{\mu}$ vanishes,
\begin{eqnarray}
\widetilde{\nabla}_{\alpha}\left(\lambda \omega^{\alpha\mu \dots
\nu}\right)=\lambda 
\left(\nabla_{\alpha}\Omega^{\alpha\mu' \dots \nu'
}\right)\Pi^{\mu}_{\ \mu'}\dots \Pi^{\nu}_{\ \nu'}\,, 
\label{formlemma}
\end{eqnarray}
where $\omega$ is the $\Sigma$-projection of $\Omega\,.$
%, and $\pounds_{\zeta}\Omega=0$. 
This is essentially the
statement that the 4-divergence of $\Omega$ is the same as its
3-divergence when both $\Omega$ and the metric are time
independent. All our proofs will be based on this result.

Let us start with the example of a real scalar field $\phi$ in a
potential $V(\phi)$.  The equation of motion for $\phi$ is
\begin{eqnarray}
\nabla_{\mu}\nabla^{\mu}\phi=\frac{\partial V(\phi)}{\partial
\phi}=V'(\phi)\,.
\end{eqnarray}
A non-vanishing $V(\phi)$ enters the stress-energy tensor, so it
follows that $\pounds_{\zeta} \phi=0$ on $\Sigma$. Then we can
project this equation down to $\Sigma$ using Eq.~(\ref{formlemma})
to get
\begin{eqnarray}
\widetilde{\nabla}_{\mu}(\lambda\widetilde{\nabla}^{\mu}\phi) =
\lambda  V'(\phi)\,.
\end{eqnarray}
For $ V(\phi)$ convex (i.e., $V''(\phi)\geq 0$), we
multiply both sides of this equation by $V'(\phi)$ and integrate
over the spacelike region $\Sigma$ between the two horizons to get
\begin{eqnarray}
&& \int_{\partial \Sigma}\lambda
V'(\phi)n^\mu\widetilde{\nabla}_\mu\phi \nonumber \\
&& \qquad\qquad + \int_{\Sigma}\lambda\left(V''(\phi) 
\widetilde{\nabla}^{\mu}\phi\widetilde{\nabla}_{\mu}\phi
+ V'^2(\phi)\right)=0\,.\quad
\label{integrals}
\end{eqnarray}
%
%
%%% \begin{eqnarray}
%%% \int_{\partial \Sigma}\lambda
%%% V' n^\mu\widetilde{\nabla}_\mu\phi +
%%% \int_{\Sigma}\lambda\left(V''\,
%%% \widetilde{\nabla}^{\mu}\phi\widetilde{\nabla}_{\mu}\phi
%%% + V'^2\right)=0
%%% \label{integrals}
%%% \end{eqnarray}
%
Here $\partial\Sigma$ is composed of the two spheres located at the
two horizons, and $n^{\mu}$ is the $\Sigma$-ward pointing spacelike
unit normal to these two spheres. Since
$\widetilde{\nabla}^{\alpha}\phi \widetilde{\nabla}_{\alpha}\phi$
appears in the stress-energy tensor $T_{\mu\nu}$, it must be
bounded at the two horizons. We may then apply Schwarz inequality,
which in this case says that
\begin{eqnarray}
|n^\alpha\widetilde{\nabla}_{\alpha}\phi|^{2}\leq
(n^\mu n_\mu) (\widetilde{\nabla}^{\alpha}\phi
\widetilde{\nabla}_{\alpha}\phi) 
 =(\widetilde{\nabla}^{\alpha}\phi
\widetilde{\nabla}_{\alpha}\phi)\,. 
\end{eqnarray}
For generic $V(\phi)$, the boundedness of $T_{\mu\nu}$ on
$\partial\Sigma$ implies that $\phi$ must also be bounded there.
Since $\lambda(r)=0$ on $\partial\Sigma$, it follows that the
integral on $\partial\Sigma$ vanishes.  Since $\Sigma$ is spacelike
$\widetilde{\nabla}^{\alpha}\phi \widetilde{\nabla}_{\alpha}\phi$
is non-negative, as is $V''(\phi)$ due to convexity, so
Eq.~(\ref{integrals}) says that $\phi$ is a constant at its minimum
everywhere on $\Sigma\,,$ which is the no hair result. For a
massless $\phi\,,$ we can multiply the field equation by $\phi$ and
insist that $\phi$ be measurable at the horizons, and the no hair
result follows. Note that we did not need to use the gravitational
equations of motion.

The proof may not apply for a non-convex potential $V(\phi)$. A
real scalar field moving in the double well potential
$V(\phi)=\frac{\alpha}{4}(\phi^2-v^2)^2$ can have a non-trivial
static solution in $\Sigma$ (it may be an unstable solution,
see~\cite{Torii:1998ir}). An interesting and not so obvious case is
that of the conformal scalar, for which the interaction is $V(\phi)
= \frac{1}{12}R \phi^2$. Then the part of the action containing
$\phi$ is invariant under local conformal transformations, as are
the scalars $T_\mu^\mu$ and $T_{\mu\nu}T^{\mu\nu}\,.$ Then in
principle one can make a transformation to make $\phi$ or
$\grad_\mu\phi$ diverge at $\partial\Sigma$ without causing a
curvature singularity. Then the $\partial\Sigma$ integral can be
non-zero, which allows a non-trivial configuration of $\phi$ on
$\Sigma\,.$ Indeed solutions with conformal scalar hair with
$\Lambda>0$ are known~\cite{Martinez:2002ru}. The proof also will
not apply to scalars with a kinetic term of the wrong sign, as in
phantom models of dark energy~\cite{Izquierdo:2005ku}. Of course,
in such models a static black hole may not form in the first place,
and a statement of no hair theorems may not be possible.

For the massive vector field, the proof proceeds in a similar
manner.  The matter Lagrangian is
\begin{eqnarray}
\mathcal{L} = -\frac14 F_{\mu\nu}F^{\mu\nu} - \frac12 m^2
A_{\mu}A^{\mu}\,. 
\end{eqnarray}
Let us define the electric potential $\phi =
\lambda^{-1}\zeta_{\mu} A^{\mu}$ and electric field $e^{\mu} = 
\lambda^{-1}\zeta_{\nu}F^{\mu\nu}$. A little algebra shows that
\begin{eqnarray}
\widetilde{\nabla}_{\mu}(\lambda \phi) = \lambda e_\mu +
\pounds_\zeta a_\mu\,, \qquad \widetilde{\nabla}_{\mu}e^{\mu} =
\lambda^{-1} \zeta_{\alpha}\nabla_{\mu}F^{\mu\alpha}\,,
\label{elecdef}
\end{eqnarray}
so that the equation of motion for $e^\mu$ is
\begin{eqnarray}
\widetilde{\nabla}_{\mu}e^{\mu}-m^2\phi=0\,.
\end{eqnarray}
Multiplying both sides by $\lambda\phi$ and integrating, we
find
\begin{eqnarray}
\int_{\partial \Sigma}\lambda \phi e^{\mu}n_{\mu} + \int_\Sigma
(\lambda(e_\mu e^\mu + m^2\phi^2) + e^\mu\pounds_\zeta A_\mu) =
0\,,
\end{eqnarray}
where $n^\mu$ is the $\Sigma$-ward unit normal to $\partial
\Sigma$, as before. Since $\phi^2$ and $e_\mu e^\mu$ both appear in
$T_{\mu\nu}$, $\phi$ must be finite, and by Schwarz inequality
$e^\mu n_\mu$ is finite, so the $\partial\Sigma$ integral vanishes.
The Lie derivative vanishes by staticity, so the vanishing $\Sigma$
integral contains positive definite quantities. It follows that
$\phi = 0 = e_\mu\,$ on $\Sigma\,.$

The equation of motion for the magnetic field is
\begin{eqnarray}
\widetilde{\nabla}_{\mu}\left(\lambda f^{\mu\nu}\right) - m^2
\lambda a^{\nu} = 0\,,
\label{procamag}
\end{eqnarray}
where $a_\mu$ and $f_{\mu\nu}$ are the $\Sigma$-projections of
$A_\mu$ and $F_{\mu\nu}\,.$ Multiplying both sides by $a_{\nu}$ and
integrating, we find
\begin{eqnarray}
\int_{\partial \Sigma} \lambda
a_{\nu}f^{\mu\nu}n_{\mu}+\int_{\Sigma}\lambda\left(\frac{1}{2}
(f^{\mu\nu})^2+m^2 (a^{\mu})^2\right) = 0\,. 
\end{eqnarray}
Since $a^{\mu}$ and $f_{\mu\nu}$ appear in $ T_{\mu\nu}$, these
must be regular, which ensures that the $\partial\Sigma$ integral
vanishes. The second integral is over a sum of squares, so $a_\mu =
0 = f_{\mu\nu}$ on $\Sigma\,,$ which is the desired no hair result.

For the massless vector field the Lagrangian has a local gauge
symmetry, which nullifies the boundedness argument. A gauge
transformation can always change a bounded function $\phi$ to one
that becomes unbounded on the horizon. Thus we cannot set the
$\partial\Sigma$ integration to zero, so $e_\mu$ need not vanish on
$\Sigma$ either. In fact Reissner-N\"{o}rdstrom solutions with a
positive cosmological constant are known.

There are two gauge-invariant Lagrangians which describe a massive
Abelian gauge field. The no-hair conjecture fails for both of these
cases in the presence of a positive $\Lambda$, as we describe now.

The first mechanism we consider is described by the Lagrangian
\begin{eqnarray}
\mathcal{L}= -\frac14 F_{\mu\nu}F^{\mu\nu} -
\frac{1}{12}H_{\mu\nu\rho}H^{\mu\nu\rho}+\frac m4
\epsilon^{\mu\nu\rho\sigma}B_{\mu\nu}F_{\rho\sigma} \,,
\end{eqnarray}
where $B_{\mu\nu}$ is an antisymmetric tensor potential and
$H_{\mu\nu\rho}= (\nabla_\mu B_{\nu\rho} + cyclic)$ is its field
strength. This system describes equally well either a massive
vector or a massive antisymmetric tensor. A static, spherical,
asymptotically flat black hole can carry a charge of the $B$ field,
with both $F_{\mu\nu}$ and $H_{\mu\nu\rho}$ vanishing everywhere
outside the black hole horizon~\cite{Allen:1990kc}. It is easy to
see that a similar solution exists for $\Lambda>0$ as well.

%%% The equations of motion are
%%% %
%%% \begin{eqnarray}
%%% \nabla_{\nu}F^{\mu\nu}  &=& \frac m6
%%% \epsilon^{\mu\nu\rho\sigma}  
%%% H_{\nu\rho\sigma}\,,\label{Feom}\\
%%% \nabla_{\sigma}H^{\mu\nu\sigma}  &=& - \frac m2
%%% \epsilon^{\mu\nu\rho\sigma} F_{\rho\sigma}\,. \label{Heom} 
%%% \end{eqnarray}
%%% %
Let $f^{\mu\nu}$ and $h^{\mu}$ be the $\Sigma$-projections of
$F^{\mu\nu}$ and $H^{\mu} \equiv \frac16
\epsilon^{\mu\nu\lambda\rho} H_{\nu\lambda\rho}\,,$ respectively.
Then the `magnetic equations' can be written as
\begin{eqnarray}
\widetilde{\nabla}_{\nu}(\lambda f^{\mu\nu})= \lambda m
h^{\mu}\,,\label{fheom} \qquad %\\*
\widetilde{\nabla}_{[\alpha} h_{\beta]} =
- m f_{\alpha\beta}\,. %\label{dualheom}
\end{eqnarray}
If we also define  $e^{\mu} = \lambda^{-1}\zeta_{\nu}F^{\mu\nu}$
and $\psi = \lambda^{-1}\zeta_\rho H^\rho\,,$ we find the `electric
equations' 
\begin{eqnarray}
\grad_\mu e^\mu = - m\psi\,, \label{eheom} \qquad % \\
\grad_\mu (\lambda\psi) = - \lambda m e_\mu\,,
%\label{dualeheom} 
\end{eqnarray}
where we have used  $\pounds_\zeta H^\mu = 0\,.$

Multiplying the first of Eq.~(\ref{fheom}) by $h_\mu$, and the
first of Eq.~(\ref{eheom}) by $\lambda\psi$, and integrating, we
obtain
\begin{eqnarray}
&&\int_{\partial\Sigma} \lambda f^{\mu\nu} h_{\mu} n_{\nu} 
+\int_{\Sigma} m\lambda \Big( \frac12 f^{\mu\nu}f_{\mu\nu} + 
h^{\mu}h_{\mu}  \Big) = 0\,,\qquad \\
&& \int_{\partial\Sigma} \lambda \psi e_\mu n^\mu - \int_{\Sigma} 
m\lambda \Big( e_\mu e^\mu + \psi^2 \Big) = 0\,.
\end{eqnarray}
The surface integrals contribute nothing. It follows that all
components of the field strengths $H_{\mu\nu\lambda}$ and
$F_{\mu\nu}$ vanish on $\Sigma$. The solution is then the de
Sitter-Schwarzschild black hole, with an arbitrary charge $q$
corresponding to the $B$-field, whose non-vanishing component is
\begin{eqnarray}
B_{\theta\phi} = \frac{q}{4\pi r^2}\,.
%\label{}
\end{eqnarray}
This charge should be measurable via a stringy Bohm-Aharonov
effect, just as for asymptotically flat
space-times~\cite{Bowick:1988xh}. We should mention here that the
free Abelian two-form will leave the same kind of charge on the
black hole, the proof of $H_{\mu\nu\rho} = 0$ on $\Sigma$ proceeds
in a similar fashion for that theory.

The other case is that of the Abelian Higgs model. In the absence
of cosmological constant, a static spherically symmetric black hole
does not carry electric (or magnetic) charge if the gauge field
becomes massive via spontaneous symmetry breaking. However, as we
shall see now, the presence of a positive cosmological constant
allows a charged black hole in the false vacuum. The matter
Lagrangian for the Abelian Higgs model is
\begin{eqnarray}
\mathcal{L}&=& - \frac14 F_{\mu\nu}F^{\mu\nu} - \frac12
q^2\rho^2(A_\mu  + \frac{1}{qv}\nabla_\mu\eta)(A^\mu +
\frac{1}{qv}\nabla^\mu\eta) \nonumber \\ 
&&\qquad\qquad - \frac12 \nabla_{\mu}\rho
\nabla^{\mu}\rho+ \frac{\alpha}{4}(\rho^2 -v^2)^2\,.
\end{eqnarray}

The equations for the magnetic and the electric fields on $\Sigma$
read 
\begin{eqnarray}
\grad_\nu(\lambda f^{\mu\nu}) + \lambda q^2 \rho^2
(a^\mu + \frac{1}{qv}\grad^\mu\eta) &=& 0\,, \label{mageq}\\
\grad_\mu e^\mu - q^2 \rho^2 (\phi + \frac1{qv}\pounds_\zeta \eta)
&=& 0\,,  
\label{eleceq}
\end{eqnarray}
where the definitions for $e^\mu$ and $f_{\mu\nu}$ are as in
Eqs.~(\ref{elecdef}) and~(\ref{procamag})\,. Applying the now
familiar techniques to Eq.~(\ref{mageq}), we get
\begin{eqnarray}
\int_{\partial \Sigma}\lambda (a_\mu &+&
\frac{1}{qv}\grad_\mu\eta)f^{\mu\nu}n_\nu 
- \int_\Sigma \lambda\left(\frac12 f^{\mu\nu}f_{\mu\nu}
\right. \nonumber \\
&+& \left. q^2\rho^2(a_\mu + \frac{1}{qv}\grad_\mu\eta)(a^\mu +
\frac{1}{qv}\grad^\mu\eta)\right)=0\,. \nonumber \\  
\label{monopole}
\end{eqnarray}
The $\Sigma$ integral can be non-vanishing only if the
$\partial\Sigma$ integral is also non-vanishing, which means that
the norm of either $f_{\mu\nu}$ or $(a_\mu + \grad_\mu\eta)$ must
diverge at the horizon. However, since we have assumed spherical
symmetry, a non-vanishing $f_{\mu\nu}$ is essentially that of the
magnetic monopole. But then $(a_\mu + \grad_\mu \eta)$ cannot be
both spherically symmetric and divergent at the horizon. So
$f_{\mu\nu} =0$ on $\Sigma\,.$

For the electric field we use Eq.~(\ref{eleceq}) to find
\begin{eqnarray}
&& \int_{\partial\Sigma}\lambda\left(\phi +
\frac1{\lambda qv}\dot{\eta}\right)e^{\mu}n_{\mu} \nonumber 
\\ && \qquad
+ \int_{\Sigma}\left[\lambda e^{\mu}e_{\mu} +
\lambda q^2\rho^2\left(\phi +
\frac1{\lambda qv}\dot{\eta}\right)^2\right]=0\,, \qquad
\label{eleceom}
\end{eqnarray}
where $\dot\eta = \pounds_\zeta\eta$, and we have used
$\pounds_\zeta(a_\mu + \frac{1}{qv}\grad_\mu\eta) = 0\,$ because of
staticity. Since $e_\mu e^\mu$ appears in $T_{\mu\nu}\,,$ we can
use Schwarz inequality to say that $e^{\mu}n_{\mu}$ is finite on
$\partial\Sigma\,.$ So the $\Sigma$ integral can be non-zero only
if $(\phi + \frac{1}{qv}\lambda^{-1}\dot\eta)$ diverges on at least
one horizon. In this case $\rho$ must vanish on that horizon.

For the asymptotically flat black hole ($\Lambda = 0\,$), it can be
shown that $\rho$ cannot vanish on the horizon, and so the black
hole cannot have electric charge~\cite{Lahiri:1993vg}. Let us see
what happens for our present choice of $\Lambda>0\,.$ The equation
of motion for $\rho$ projected down to $\Sigma$ reads
\begin{equation}
\grad_{\mu}\lambda \grad^{\mu}\rho = - \lambda q^2\rho\left(\phi + 
\frac{1}{qv}\lambda^{-1}\dot{\eta}\right)^2 +  \lambda\alpha\rho
(\rho^2-v^2).   
\label{rhoeq}
\end{equation}   
Let us assume for the moment that $\rho$ vanishes on the black hole
horizon at $r = r_H$, and starts increasing with increasing $r\,.$
Then $\rho$ must increase monotonically from $\rho=0$ at $r=r_H$ to
one of: $(i)\, \rho = \rho_C < v$ at $r = r_C\,;$ $(ii)\, \rho = v$
at $r = r_v \leq r_C\,;$ $(iii)\, \rho = \rho_{max} < v$ at the
turning point $r = r_{max} < r_C\,.$

In all three cases, we multiply Eq.~(\ref{rhoeq}) by $(\rho - v)$
and integrate over a region $\Omega\,$ to get 
\begin{eqnarray}
&&\!\!\int_{\partial\Omega}\lambda(\rho - v)n^\mu\grad_\mu\rho -
\int_\Omega\lambda \left[\grad_\mu\rho\grad^\mu\rho \right.
\nonumber \\ && \left.- \rho(\rho-v)\left(\phi + \frac{1}{\lambda
qv} \dot\eta\right)^2 +  \alpha(\rho-v)^2\rho(\rho+v)\right]=0
\,.\nonumber \\
\end{eqnarray}
The region $\Omega$ and its boundary $\partial\Omega$ for the three
cases are taken respectively to be $(i)\, \Omega =
\Sigma\,,\partial\Omega = \partial\Sigma\,;$ $(ii)\, \Omega
=\Sigma\vert_{r<r_v}\,,$ $\partial\Omega =$ spheres at $r_H,
r_v\,;$   $(iii)\, \Omega = \Sigma\vert_{r < r_{max}}\,, $ 
$\partial\Omega =$ spheres at $r_H, r_{max}\,.$

In all three cases, the integral over $\partial\Omega$ vanishes,
and all terms in the $\Omega$ integral are non-negative everywhere
on $\Omega\,.$ So we have a contradiction and $\rho$ cannot
increase from zero as $r$ increases from $r_H\,.$ These arguments
can be trivially modified to show that $\rho$ cannot decrease from
zero as $r$ increases from $r_H\,,$ nor can $\rho$ increase or
decrease from zero as $r$ decreases from $r_C\,.$ So in general,
$\rho\neq 0$ at either horizon, so the electric field vanishes on
$\Sigma\,,$ and the black hole does not carry an electric charge,
which is the no-hair statement. There is however one exception.
This is the solution for which $\rho = 0$ on all of $\Sigma\,.$
Then Eq.s~(\ref{eleceq}) and ~(\ref{eleceom}) are the same as those
for the ordinary Maxwell-Einstein system. Then the black hole may
carry an electric charge, and the space-time is described by the
Reissner-N\"ordstrom-de Sitter solution.

We also note here that the assumption of spherical symmetry is not
crucial for the proofs, except for the Higgs model. So axisymmetric
black holes are hairless for most field theories, while dipole or
other axisymmetric hair cannot be ruled out for the Higgs model.

We have proved various no-hair theorems by restricting attention to
the region between the two horizons for black hole space-times with
$\Lambda>0\,.$ Unlike usual investigations of black hole
space-times, we have managed to completely ignore the asymptotic
behavior. This is the new paradigm referred to earlier, which we
believe should be useful in further investigations of $\Lambda>0$
space-times.  Interestingly, the Abelian Higgs system allows a
charged solution which has no counterpart in the asymptotically
flat case. This suggests the intriguing possibility that, even for
the $\Lambda=0$ black holes with hair, there may be additional
classes of solutions for $\Lambda>0\,,$ coming from non-trivial
boundary conditions at the two horizons.  For example, black holes
pierced by a cosmic string~\cite{Achucarro:1995nu}, black holes
with non-trivial external Yang-Mills and Higgs fields, or Skyrme
black holes~\cite{Volkov:1998cc,Brihaye:2006kn}\, may have more
varied counterparts for $\Lambda>0\,.$ Black holes with discrete
gauge hair (see~\cite{Coleman:1991ku} for a review), because of the
underlying Higgs model, may be dressed differently for
$\Lambda>0\,.$ There may also be new axisymmetric solutions in a
Higgs background. Other kinds of quantum hair such as non-Abelian
quantum hair~\cite{Coleman:1991ku, Lahiri:1992yz} or spin-two
hair~\cite{Dvali:2006az}, whose existence are related to the
topology of the space-time, are likely to be present also for
$\Lambda>0\,.$

%\bigskip
%{\bf References}

%%%%%%%%%%%%%%%%%%%%%%%%%

\end{document}